\documentclass[prb,twocolumn,showpacs,amsmath,amssymb,floatfix]{revtex4}
\usepackage{graphicx}
\usepackage{dcolumn} 
\usepackage{bm}
\usepackage{amsmath}
\usepackage{latexsym}
\usepackage{amsfonts}
\usepackage{amssymb}
\usepackage[rightcaption]{sidecap}
\newcommand{\be}{\begin{equation}}
\newcommand{\ee}{\end{equation}}
\newcommand{\bea}{\begin{eqnarray}}
\newcommand{\eea}{\end{eqnarray}}
\newcommand{\nn} {\nonumber}
\renewcommand{\vr} {{\bf r}}

\newcommand{\ro} {{\rm o}}

\newcommand{\Tr}{ {\rm Tr} \, }

\def\a{\alpha}

\def\l{\lambda}

\def\S{\Sigma}

\def\F{\Phi}

\def\w{\omega}

\def\xc{{\rm xc}}
\def\H{{\rm H}}
\def\Tr{{\rm Tr}\,}

\def\br{\mbox{\boldmath $r$}}

\begin{document}
\preprint{APS/123-QED}
\title{The correlation potential in density functional theory at the GW-level: \\
spherical atoms}
\date{\today}
\author{M. Hellgren}
\author{U. von Barth}
\affiliation{Solid State Theory, Institute of Physics, Lund University, 
S\"olvegatan 14A, S-22362 Lund, Sweden}  
\date{\today}              
\begin{abstract}
As part of a project to obtain better optical response functions for
nano materials and other systems with strong excitonic effects we
here calculate the exchange-correlation (XC) potential of density-functional
theory (DFT) at a level of approximation which corresponds to the dynamically-
screened-exchange or GW approximation. In this process we have designed
a new numerical method based on cubic splines which appears to
be superior to other techniques previously applied to the "inverse
engineering problem" of DFT, {\em i.e.}, the problem of finding an XC
potential from a known particle density. The potentials we
obtain do not suffer from unphysical ripple and have, to within a
reasonable accuracy, the correct asymptotic tails outside localized
systems. The XC potential is an important ingredient in finding the
particle-conserving excitation energies in atoms and molecules and
our potentials perform better in this regard as compared to the LDA
potential, potentials from GGA:s, and a DFT potential based on
MP2 theory.
\end{abstract}
\maketitle
\section{Introduction}
The work presented here is part of an ongoing investigation aimed at
finding more accurate and computationally efficient ways to calculate
optical spectra from materials with strong excitonic effects. Typical
examples are most semiconductors, rare-gas crystals, atoms and molecules
and nano materials. In these systems the attraction of the excited
electron to the hole created in the excitation process is very important
and will strongly influence the measured spectra both in the continuum
and in the discrete part of the spectra. This so called particle-hole
effect or vertex correction must be
accounted for in the theoretical description in order to have a
reasonably accurate, quantitative agreement with the experimental
results.

The traditional way to incorporate the particle-hole attraction within
many-body perturbation theory is to obtain the optical spectra from
solutions to the Bethe-Salpeter equation for the four-point vertex
function. Unfortunately, this is a very demanding procedure already in 
high symmetry cases like crystalline solids and the method becomes very
tough indeed when the symmetry is much lower like, {\em e.g.}, at surfaces or,
even worse, in nano systems. In solids, the two-body wave function
describing the particle and the hole is usually expanded in valence-
and conduction-band one-electron orbitals leading to huge secular
problems.

Nevertheless, the Bethe-Salpeter approach has provided a large number of
accurate and useful results in many different systems in spite of the
fact that many approximations enter the theory like, {\em e.g.}, that the
actual particle-hole interaction is most often replaced by a statically
screened potential. During the past ten years, Time-Dependent
Density-Functional Theory (TDDFT) has emerged as a competing and,
perhaps, more efficient method for calculating optical absorption
spectra including excitonic effects. The central quantity of this approach
is the so called exchange-correlation (XC) kernel the knowledge of
which, unfortunately, is rather limited. The kernel is the functional
derivative with respect to the density of the XC
potential which is also a relatively unknown quantity. Due to the
pioneering work of several groups within some of the research networks
of the European Union\cite{revlucia,botti} it has been shown
that the accurate results of the Bethe-Salpeter approach can be
reproduced within the framework of TDDFT provided an appropriate XC
kernel is used in the calculation. These results are very encouraging,
the drawback being that the method does not provide a recipe for
improving the kernel without first improving the underlying
approximations within the Bethe-Salpeter approach. But there is a
great computational advantage of any approach based on TDDFT as
compared to a more standard many-body approach like, {\em e.g.}, that based
on the Bethe-Salpeter equation. The former theories are based on
two-point correlation functions whereas the latter on four-point
correlation functions.

The simplest approximation to the XC kernel is the so called adiabatic
local-density approximation (ALDA) in which the potential is taken to be
the exchange-correlation part of the chemical potential of the electron
gas evaluated at the instantaneous electron density. This simple
approach is, however, known to suffer from many ailments. The
accurate results of the Bethe-Salpeter approach can, {\em e.g.}, not be
reproduced. The next level of approximation is the exchange-only (EXO)
approximation which, to our knowledge, has only been applied once to
solids,\cite{kg} unfortunately, with rather poor results as far
as the spectral properties are concerned. The approximation seems,
however, to give reasonable excitation energies in localized systems and
the total energies calculated from the corresponding response function
are rather accurate in all systems.\cite{pgg,kvB}

Until recently, one of the drawbacks of TDDFT has been the lack of a
systematic approach for obtaining successively better approximations to
the XC kernel. Straight-forward many-body perturbation theory (MBPT)
can certainly be used to generate approximations to the electronic self
energy and the three-point vertex function. But the subsequent conversion
into an XC kernel of TDDFT is not at all guaranteed to yield, {\em e.g.},
a particle conserving density response function.\cite{punk,lucia}
Particle and current conservation are important properties of
physical response functions which should be built-in to
their construction. A systematic theory for improved kernels within
TDDFT has recently been introduced by us.\cite{tddftvar} This
theory automatically leads to conserving response functions. It is based
on an adaptation to TDDFT of our variational approach to many-body
perturbation theory.\cite{abl} A variational functional of the one-electron
Green function is constructed which yields the total energy of the
system when evaluated at that Green function which renders the
functional stationary. Because of the stationary property of the
functional and thus an absence of first-order errors, accurate energies
can be obtained already at rather crude Green functions like, {\em e.g.}, a
non-interacting one.\cite{nedvbatm} The construction of the functional
has two basic ingredients: i) a choice of basic functional expression
ultimately responsible for the variational quality of the total
functional (the size of the second-order corrections to the energies).
So far, only two such basic expressions have been considered, one due
to Luttinger and Ward\cite{lw} and a simpler one due to Klein.\cite{klein} The former was shown to have better variational properties than the latter.\cite{nedvbatm}
However, as we have shown previously,\cite{tddftvar} the LW functional leads 
to a response function the calculation of which is beyond our 
present day capabilities in realistic systems. Therefore,
we will here consider only the Klein version of the functional.
ii) The choice
of $\Phi$ functional which is also a functional of the one-electron
Green function $G$.
The formal significance of the $\Phi$-functional is that its functional
derivative with respect to $G$ yields the electronic self-energy and
its physical significance is that it will contain our physical intuition
concerning the importance of different physical processes. For
instance, in an extended system all the screening diagrams should be
included in the  $\Phi$-functional and in a system with strong
correlation effects it would be appropriate to include some
particle-hole and particle-particle ladders. In the present study we
will focus on the screening diagrams in the
$\Phi$-functional and the
resulting electronic self-energy will thus be that of the GW
approximation.\cite{hedin} Some consideration will, however, be
given also to the second-order exchange diagram.
The variational approach to MBPT is converted into a
density-functional theory by a restriction of the variational freedom
for the Green function which is taken to be one pertinent to
a non-interacting system moving in some local external potential
$V(\vr)$. Due to the Hohenberg-Kohn theorem applied to non-interacting
electrons and the one-to-one correspondence between the applied
potential and the particle density this restriction immediately turns
the variational functional of the Green function into a variational
functional of the density. A rewriting of the Klein functional in
terms of the particle density recovers the normal form of the total
energy of DFT\cite{KS} in which the so called XC energy becomes the
$\Phi$ part of the functional. Functional differentiation with respect
to the density yields the XC potential which is the central object of
interest in the present work.

Worked aimed at finding the correct density-functional (DF) potential
pertinent to different approximations started a long time ago. Even
before the advent of DFT, Sharp and Horton\cite{sh} proposed a method for
finding a local potential which would accurately reproduce the total
energies and densities of atoms within the Hartree-Fock approximation.
This work is, in todays language, best referred to as the first
appearance of the Exchange-Only (EXO) approximation or the Exact-Exchange (EXX) method. 
This method, sometimes also referred to as an optimized potential method (OPM)
was later used by Talman and Shadwick\cite{ts} for doing calculations on
many atoms. In 1982, one of us made use of the Hohenberg-Kohn theorem to
find that local potential which exactly reproduces the Hartree-Fock
densities of several atoms.\cite{rodsuhl} These calculations are not equivalent to the
EXO but the results are, numerically, very close indeed. Shortly
afterwards, this numerical fitting procedure was generalized to include
also all correlation effects thus producing\cite{cp} the first
exact DF XC potentials for several atoms.
As a matter of fact, the exact XC potential for
the Helium atom was known already at that time through the work of
Smith, Jagannathan and Handler.\cite{sjh}
Aryasetiawan and Stott\cite{awst1,awst2} also found the exact XC potential of DFT for
some smaller atoms using a presumably more
accurate approach which had the additional advantage of offering insight
into the so called v-representability problem of DFT. In solids, early
progress toward an exact DF XC potential was made by Godby {\em et al.}\cite{godbyGW1,godbyGW2} who actually constructed their XC potential as a solution to
the Linearized Sham-Schl\"uter equation (LSS).\cite{godbyGW2} The self-energy of 
their choice was again that of the GW approximation (GWA). Through the formal 
proofs of the present work we now know that their potential was in fact the full 
RPA XC potential of DFT for the semiconductors they studied. Consequently,
the work of Godby {\em et al.} was almost identical in spirit to that
of the present work, albeit in solids.\cite{txt}

From the middle of the eighties the number of publications describing
work aimed at finding improved XC potentials for DF calculations
increased rapidly - both with regard to approximations for use in
practical calculations and with regard to a fundamental understanding
of the behavior of the potentials in exact cases and in model systems.
The most accurate XC potentials produced thus far are probably those
published by Umrigar and Gonze.\cite{umr}

In the present work we concentrate on the XC potential at the GW level
which we here prove to be identical to that which minimizes the standard
expression for the total energy within the Random Phase Approximation
(RPA). Our main motivation is an interest in density and current
response functions beyond the EXO and the XC potentials then constitutes
one of the basic ingredients. Through the work of, {\em e.g.}, Petersilka {\em et al.}
\cite{pgb} it has, however, long been realized that the accuracy of
the XC potentials is crucial for obtaining accurate excitation
energies from TDDFT. Thus, it is certainly of interest to see how
well our GW-based potentials perform in this context. In addition, these
potentials provide a good testing ground for our new numerical approach
based on splines as basis functions for electronic structure
calculations in atoms, molecules, and solids.

The paper is organized as follows. In Sec. II we shortly present the
formal framework based on the variational approach to MBPT.
In Sec. III we present our new numerical
approach and discuss its advantages and shortcomings.
Numerical results for several spherically symmetric atoms 
are given in Sec. IV. We discuss the behavior of the 
GW/RPA potentials and compare their performance to that of  
potentials of other approximations, like, {\em e.g.}, the EXX and the 
MP2. We also calculate particle-conserving excitation energies
using our calculated potentials in conjunction with the approximate 
XC kernel of PGG.\cite{pgg} Finally, in Sec. V, we draw our conclusions
as well as advertise our forthcoming work on response
functions.
\section{Conserving approximations within TDDFT}
Physical observables of a system of interacting electrons can be calculated within MBPT, where the central quantity is the 
one-particle propagator, or the Green function $G$. The latter has a diagrammatic expansion in powers of the Coulomb interaction which, in extended systems, always must be carried to infinite order. Guided by physical intuition, approximations for $G$ can be constructed by including only a 
selected set of diagrams, appropriate to the system studied. The expansion of $G$ can be written in terms of Dyson's equation,
\be
G=G_{\rm H}+G_{\rm H}\S G,
\label{dys}
\ee
where $G_{\H}$ is the Hartree Green function and $\S$ is the self energy which
contains all the many-body effects above the Hartree level. The Hartree Green function 
$G_{\rm H}$ is the one-electron propagator of non-interacting electrons moving in the total potential consisting of the external potential $w$ and the Hartree potential, {\em i.e.}, 
the Coulomb potential from the total electronic charge density. 

Within MBPT, also the density response function $\chi$ has an expansion in
powers of the Coulomb interaction. Choosing only a subset of diagrams, albeit 
an infinite subset, results in an approximate response function which only by 
pure chance will obey important conservation laws like, {\em e.g.}, particle number,
momentum and energy conservation. A scheme to construct approximations within 
the framework of MBPT which {\em are} conserving was first proposed by Kadanoff 
and Baym.\cite{bk,baym} They made use of a functional $\F[G]$ with the property 
that its functional derivative with respect to $G$ is the self energy,\cite{lw}
\be
\Sigma=\frac{\delta\F}{\delta G}.
\ee
As a consequence, the self energy $\S$ will be a functional of the 
interacting Green function $G$. The response to external perturbations of a 
system treated within such an approximation will involve the derivative of the
self energy with respect to $G$ and consequently a symmetric second
derivative of $\F$ with respect to $G$. It can be shown that this
symmetry is a sufficient condition for the conserving properties of the
resulting response function. Approximations generated from this scheme are
called $\F$-derivable.

About the same time, Klein\cite{klein} constructed a variational functional of $G$ composed of the $\F$-functional and some additional terms,
\be
 i Y_{\rm K}[G]=\F[G]-\Tr\left\{
GG_{\rm H}^{-1}-1+\ln(-G^{-1})\right\} - i U_{\ro}[G].
\ee
Here, $U_\ro$ is the classical Coulomb interaction energy between the 
electrons given by $U_\ro=\frac{1}{2}\int nvn$. When this Klein functional 
is varied with respect to $G$ it is seen to be stationary 
when $G$ obeys Dyson's equation, Eq. (\ref{dys}). Moreover, at the stationary point the functional takes the value of the ground state energy of the system. The functional is general and applies to any system; the reference to the 
particular system is contained in the Hartree Green function, $G_{\rm H}$.
Other functionals of $G$ and $\F$ have been designed like, {\em e.g.},
the LW functional\cite{lw} or the ABL functional.\cite{abl} These functionals have different and generally better variational properties as compared to the Klein
functional and we refer the reader to Ref. \onlinecite{nedvbatm} for a
more comprehensive discussion. In this work, however, we will
focus our attention on the Klein functional.

Starting from the Klein functional, we can restrict the variational
freedom of the Green function to non-interacting ones, $G_s$,
generated by a local multiplicative potential, $V$. The Klein
functional then becomes a functional of that potential. From the
Hohenberg-Kohn theorem there is a one-to-one mapping between the
particle density and the potential which turns the Klein
functional into a functional of the density and our theory
into a time dependent density functional theory.
The simplicity of a non-interacting Green function allows us to
convert the Klein functional into the form,
\be
Y_{\rm K}[V]= -i \F[G_{s}] + T_s[n] + \int w n
 + U_\ro,
\label{energi}
\ee
where $T_s$ is the kinetic energy of non-interacting electrons and 
$w$ is the external potential. It is now clear that $-i\F$ plays the 
role of the XC energy, $E_{\xc}$.\cite{KS} Varying this form with respect to the potential $V$ we find it to be stationary when the potential is given by
$$
V=w+V_{\rm H}+v_{\xc},
$$
where $V_{\rm H}=\int nv$ is the Hartree potential and
where $v_{\xc}$ obeys
\begin{equation}
\int i\chi_{s}(1,2)v_{\xc}(2) \; d2 =
\int \S(2,3)G_s (3,1)G_s(1,2) \; d(23).
\label{LSS}
\end{equation}
This is the well known so-called linearized Sham-Schl\"uter (LSS) equation\cite{godbyGW2} 
which here is seen to follow from a variational
principle rather than being just the first iteration of an infinite number
of iterations leading to the solution to the full Sham-Schl\"uter equation.
\cite{sonly,ssb} We here remark that there is also a self-consistency procedure 
involved in solving the LSS because the non-interacting so-called Kohn-Sham 
(KS) response function $\chi_s$ as well as the self energy $\S$ are both expressed in orbitals obtained from solving a one-electron Schr\"odinger equation in which the unknown XC potential $v_\xc$ is part of the local potential. As pointed out by Casida, it is also worth noting that $v_\xc$ 
obtained in this way can be seen as the best local approximation to the self energy in a variational sense.\cite{casida} 

The conserving properties, and particle conservation in particular, are
important when calculating response functions from TDDFT. Within TDDFT,
the interacting density response function $\chi$ can be shown to be given by\cite{hardyfxc}
\be
\chi=\chi_s+\chi_s[v+f_{\xc}]\chi,
\label{fxc}
\ee
where
$$
f_{\xc}=\frac{\delta v_{\xc}}{\delta n}.
$$
A further variation of Eq. \ref{LSS} with respect to the density 
gives us an equation for the two-point XC kernel, $f_{\xc}$.\cite{tddftvar} 
And, because of the underlying $\F$-derivability of the theory the resulting response function obeys, {\em e.g.}, particle conservation which, in the linear
response regime, amounts to the {\em f-}sum rule.
Another important property which also follows from the variational and
conserving character of this theory is the well known Virial Theorem
which can be used as a stringent test of the accuracy of the calculation
of the total energy. In Appendix A the reader can find an explicit
derivation of that theorem in the context of the present theory.
\begin{figure}
\begin{flushleft}
\resizebox{!}{1.05cm}{\includegraphics{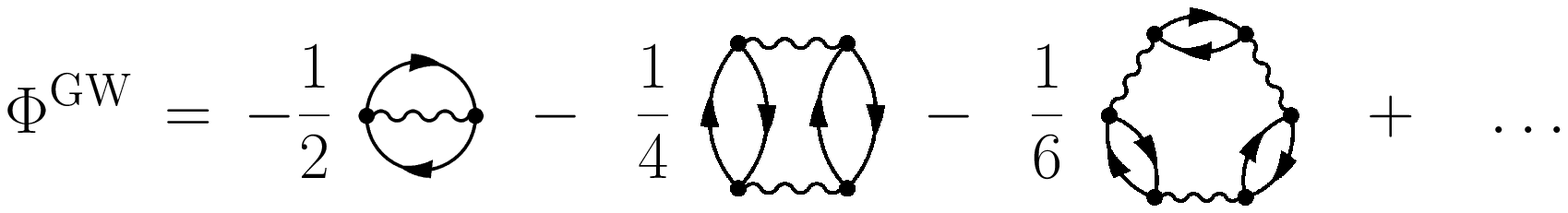}}\\
\vspace{0.7cm}
\resizebox{!}{1.05cm}{\includegraphics{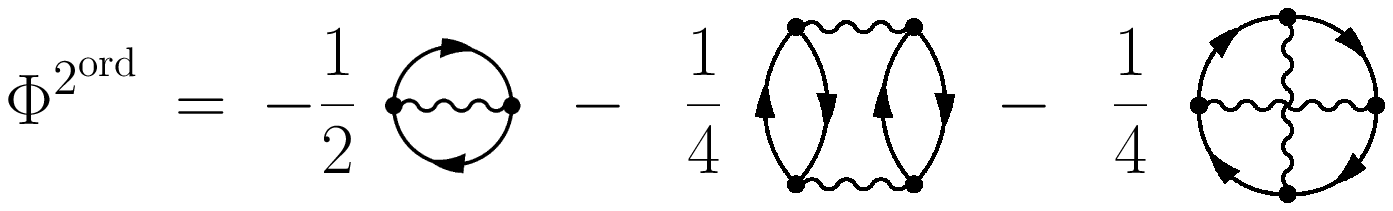}}
\end{flushleft}
\caption{Diagrammatic representation of the $\F$-functional in the two 
approximations studied in this paper.}
\label{diagram}
\end{figure}
\subsection{The GW approximation}
In this work we are interested in investigating the DF approximation
resulting from choosing $\F$ in the GWA, see Fig. \ref{diagram}. This approximation corresponds to a summation of all ring 
diagrams, where the propagators are KS Green functions,
\be
\F^{\rm GW}=\frac{1}{2}\Tr\{\ln(1+ivG_sG_s)\}.
\label{phirpa}
\ee
Inserting $\F^{\rm GW}$ into the Klein functional results in an energy expression corresponding to the well known random phase approximation (RPA) for the total energy. Consequently, the total energy within the RPA is a variational expression and the stationary point has been proven to be a minimum.\cite{nedvlvb}
Its minimization with respect to the density yields the ground-state
energy and density as well as the corresponding XC potential $v_{\xc}$.

Taking the functional derivative of $\F$ in Eq. (\ref{phirpa}) we obtain the 
self energy,
\be
\S(1,2)=iG_s(1,2)W(1,2),
\ee
where the effective interaction $W$ is given by
\be
W=[1-v\chi_s]^{-1}v,
\ee
and $\chi_s$ is the KS non-interacting polarization propagator.
The LSS-equation can then be split into two terms and written symbolically as
\begin{equation}
\int \chi_s v_{\xc}=\int
iG_s[v+v\chi^{\rm RPA} v]G_s.
\end{equation}
Keeping only the first term, the Hartree-Fock term, results in what is known
as the exact exchange (EXX) approximation or, sometimes, the exchange only (EXO) 
approximation and has been discussed earlier by
several people.\cite{ts,pgg,kg,kvB,sh} The second term gives the correlation part
of the potential and is expressed in terms of the interacting polarization
propagator $\chi^{\rm RPA}$, which, within the ring approximation or the RPA, is given by
\be
\chi^{\rm RPA}=\chi_s+\chi_s v\chi^{\rm RPA}.
\label{RPA}
\ee
In the following we will denote the self-consistent XC potential corresponding to the GW-level of the Klein functional by $v_{\xc}^{\rm RPA}$. This potential  is more commonly referred to as the XC potential of the RPA.
\subsection{The second order approximation}
In the previous section a summation of $\F$-diagrams up to infinite order in 
the Coulomb interaction is carried out. A conserving approximation does, however, not require an infinite set of $\Phi$-diagrams. In the conserving 
second order approximation, also known as the Born approximation, all diagrams 
which are at most second order in the Coulomb interaction and only those diagrams are included in the 
$\F$-functional, see Fig \ref{diagram}. Except from the first order Fock diagram 
there are then, in total, two more diagrams to be considered. One is the 
first screening diagram, also included in the GWA. The second is the first 
vertex diagram also referred to as the second order exchange diagram and it is 
not included in the GWA. 

When considering only non-interacting KS Green functions and choosing $\F$ in 
the second order approximation the resulting energy expression is easily seen 
to be equivalent to the second order energy expression obtained in M\o ller
-Plesset perturbation theory (MP2). This approximation is common 
in quantum chemistry where the inserted orbitals are those  
of the Hartree-Fock approximation. In our variational approach also the MP2 expression for the total energy can be minimized with respect to the density thus yielding a minimizing local XC potential. This potential was recently 
calculated by Jiang and Engel.
\cite{engeljiang} For the purpose of testing our new numerical 
approach described in the next section, we have repeated their calculation. In accordance with the notation of their paper we will denote that correlation potential $v^{\rm MP2}_{\rm c}$.
\section{Numerical approach}
The solution to the linearized Sham-Schl\"uter equation is complicated by the
singularities of the kernel $\chi_s$ giving rise to numerical instabilities.
Also, $\chi_s$ contains an infinite number of unoccupied states leading to
integrals over the continuum. When, like here, a finite basis is used these
integrals turn into discrete sums. Previous atomic and molecular calculations
have lead to both unphysical oscillations and to an incorrect asymptotic
behavior of the potential, as has been discussed by many authors.
\cite{engeljiang,asympdiv,unique,unique2,soleng}
In an attempt to avoid these difficulties we have designed a new
basis set consisting of cubic splines.

\subsection{Cubic splines as radial basis functions}
We start by distributing a set of five nodes, not necessarily equidistant,
along the radial axis,
$$
R=\{r_k:k=0,\ldots ,4;r_k<r_{k+1};r_k\in\mathbb{R}\}.
$$
From these nodes we can form a localized, piecewise third-order polynomial
function $S$, in the following way:
\begin{enumerate}
\item{Cubic polynomials $P_k=a_kr^3+b_kr^2+c_k r+d_k$, are defined on the four
intervals $I_k=[r_{k-1},r_{k}]$,  $k=1,\dots,4$. In each of these $S(r)=P_k(r)$.}
\item{When $r\le r_0$ and $r\ge r_4$, the function $S$, is zero.}
\item{The function $S$, is required to be continuous and to have a continuous
first and second derivative on the \emph{whole} real axis. This means that for
$k=1,\ldots,3$:}
\bea
P_{k}(r_{k})&=&P_{k+1}(r_{k})\nn\\
P'_{k}(r_{k})&=&P'_{k+1}(r_{k})\nn\\
P''_{k}(r_{k})&=&P''_{k+1}(r_{k}),\nn
\eea
and at the end-points:
\bea
P_1(r_0)=0,&\,\, &P_4(r_4)=0\nn\\
P'_1(r_0)=0,&\,\, &P'_4(r_4)=0\nn\\
P''_1(r_0)=0,&\,\, &P''_4(r_4)=0.\nn
\eea
\end{enumerate}

Functions designed in this manner are in the numerical literature called cubic
splines.\cite{splineref} From the way the spline above is constructed there
are sixteen unknown coefficients of the four cubic polynomials and fifteen
matching conditions. The coefficients can then be determined up to a common
factor. By fixing this factor the spline is uniquely defined on the given set
$R$.

For the purpose of building up a basis set, we distribute a set of mesh
points $M=\{r_k:k=0,\ldots ,N+3;r_k<r_{k+1};r_k\in\mathbb{R}\}$, along
the radial axis. From the set $M$ we can define the subsets
$R_i=\{r_k:k=i,\ldots ,i+4;r_k\in R\}$ and on every subset $R_i$ we can
define a spline $S_i$. This generates a basis set of $N$ splines with a
distribution in space determined by the choice of mesh.
This mesh should certainly be chosen to suit the physical problem at
hand. One of the first computer codes for atomic calculations was
constructed by Herman and Skillman\cite{esc} who chose radial mesh points with a
separation increasing quadratically with the distance from the nucleus -
a so called cubic mesh.
In later years an exponential mesh became more common but the idea is
similar. It is important to stack points close to the nucleus in order
to account for the rapid oscillations produced by the strong nuclear
potential while a much coarser mesh is sufficient in the outskirts of
the atom where the wave functions decay exponentially. The exponential
mesh is extreme in the sense that one obtains a very accurate
description close to the nucleus whereas that mesh gives a poor
description some distance away from the atom. This is fine if only 
occupied states of single atoms are considered. MBPT requires also a 
reasonable description of the excited one-electron orbitals for which 
the exponential mesh is inadequate. It would also be less appropriate if 
one should like to add another atom some distance away to form a molecule. 
Thus, we have here settled on a cubic mesh and have chosen our mesh points
according to $r_k = [h(k-3)]^3, k=0,\ldots,N+3$ where the 'spacing' $h$
is determined by the relation $r_{\rm max} = [h(N-3)]^3$. This choice is
further supported by convergence tests carried out by Stankovski.\cite{martin} Consequently, our
entire numerical procedure has two basic parameters, the number of
splines $N$ and the maximum radius $r_{\rm max}$ outside which no physics is
of interest to us. We are, {\em e.g.}, not interested in highly excited
states or in scattering problems. We are interested in low lying
excitations or in higher excitations only to the extent that they
indirectly affect the low-energy excitations. Of course, the convergence
of our results with respect to both these numerical parameters have
been thoroughly checked. But we find it essential to stress that, due to
the completeness of the splines, the results must converge toward the
exact results for the chosen physical approximation when $N$ and
$r_{\rm max}$ are made arbitrarily large. Thus, there is no need to discuss
the dependence of the results on the quality of the chosen basis set and
on the choice of particular exponents of Gaussians or of Slater
functions.

The KS equation is a second order differential equation the solutions of
which are required to have a continuous first derivative. A potential
with no discontinuities in the form of steps gives rise to
solutions with a second derivative which is also continuous.
Consequently, our cubic
splines fulfill the basic requirements for radial basis functions and
they have the additional advantage of constraining the potential to be
continuous, which is a property reflecting our prejudices concerning
a proper XC potential.

Before ending this subsection, we would like to mention a further
important numerical consequence of using splines as basis functions.
Every single spline does only overlap with its three closest neighbors
on both sides. Consequently, the matrices of the corresponding secular
problem are band matrices for which there exist numerous efficient
diagonalization algorithms.

\subsection{Re-expansion procedures}

The general practical procedure followed here in order to find the
XC potential pertinent to the GW approximation is as follows. A starting
potential like that of the simple LDA is used to generate a KS 
non-interacting Green function $G_s$. From this Green function we easily
obtain the KS non-interacting density response function $\chi_s$.
The time-ordered version of $\chi_s$ is
\be
\chi_s(\vr,\vr',\w)=\sum_q \frac{f_q(\vr)f^*_q(\vr')}{\w^2-(\w_q-i\eta)^2},
\ee
where $q=(k,\mu)$ is a particle-hole index, $\omega$ is the frequency,
$\omega_q = \epsilon_{\mu} - \epsilon_k$ is a particle-hole excitation
energy,
and $f_q$ is an 'excitation amplitude', {\em i.e.}, a product of the
occupied
KS orbital $\varphi_k$ and the unoccupied KS orbital $\varphi_{\mu}$.
The fact that the KS response function $\chi_s$ is diagonal in this
'excitation basis' allows for a very simple and efficient way of solving
for the full RPA density response function $\chi^{\rm RPA}$ of Eq. (\ref{RPA}).

Consequently, a substantial part of the numerical effort must be spent
on finding an efficient an accurate way of representing the product of
two one-electron orbitals. Just solving the ordinary KS equations
presents a similar problem. At every step toward self-consistency the
electron density, being the sum of the squares of the occupied orbitals,
must be re-calculated. In a numerical approach based on localized orbitals
an obvious choice of basis for the products of the wave functions would be the
product of the basis functions. With $N$ basis functions, this means that one
would use $N^2$ basis functions for the product functions meaning,
{\em e.g.}, that the matrix describing the response function $\chi_s$ would
be $N^2 \times N^2$. This is clearly an unnecessary effort. For instance,
people using an approach based on LMTO:s would use products of LMTO:s
for describing $\chi_s$ but would only use a subset of the $N^2$ products
of LMTO:s. The actual product basis functions included can, {\em e.g.},
be determined numerically by measuring the degree of variational freedom
gained by adding one extra product basis function.\cite{awst3} Unfortunately,
in this way, one may be able to reduce the number basis functions for
the products by a factor of two or three which is not a very large gain
if $N$ is large. In the case of methods based on plane waves the
situation is slightly better. If the accuracy of the wave functions are
considered enough by including all plane waves up to a chosen momentum
cut-off, quantities like the density and the excitation functions will
contain, on the average, eight times as many plane waves as the
wave functions.

In our case, using the $N$ cubic splines as basis functions, our
excitation functions would be a sum of polynomials of degree six.
But if it is sufficient to describe the wave functions in terms of $N$
polynomials of the third degree, intuition suggests that the same should
be true also for the density and the excitation functions. We thus
re-expand products of wave functions in terms of the same set of cubic
splines as used for the wave functions. The accuracy of this intuition
has to be verified numerically. We have found that, in the case of a
fixed atomic potential - {\em i.e.} no charge density or
self-consistency is involved - the accuracy of the KS
eigenvalues increases by an order of magnitude and the relative
error decreases to $10^{-5}$ when increasing the number of splines
from twenty to thirty. However, including also errors arising from
the re-expansion of the self-consistent density in terms of the same
number of cubic splines as used for the wave functions this error
increases by a factor of three at thirty splines. By a rather minor
increase in the number of splines - as compared to eight times the
number of plane waves or the number of localized basis functions
squared, we regain the full accuracy ($10^{-5}$) at just below
forty cubic splines. From this we conclude that the cubic splines
constitute a superior basis set for many-body calculations involving
two-point correlation functions.

The so obtained functions $\chi^{\rm RPA}$ and $G_s$ are finally used to
calculate the screened interaction $W$ and the self-energy $\Sigma$
of the GWA. And then the LSS, Eq. (\ref{LSS}) is solved for the XC
potential $v_{\xc}$ by expanding it too in our cubic splines and
inverting $\chi_s$ expressed as a matrix in cubic splines.
As discussed above this matrix is singular. The physical reason
for this is that the response of a constant potential is zero.
Inverting $\chi_s$ is thus not a unique operation. This difficulty
can be circumvented by adding the constraint:
\be
\lim_{r\to\infty}v_{\xc}(r)=0.
\ee
Mathematically this means to invert the matrix $\chi_s$ in the
subspace orthogonal to that defined by the zero eigenvalue of $\chi_s$.

In all our calculations the EXX potential was first 
calculated and then used as a starting guess. 
The convergence criterion in our calculations was set to $|n^{(k)}(r)-n^{(k-1)}(r)|\leq10^{-5}$, 
and was reached in a few iterations for all spherical atoms up to Ar. 
\begin{figure}
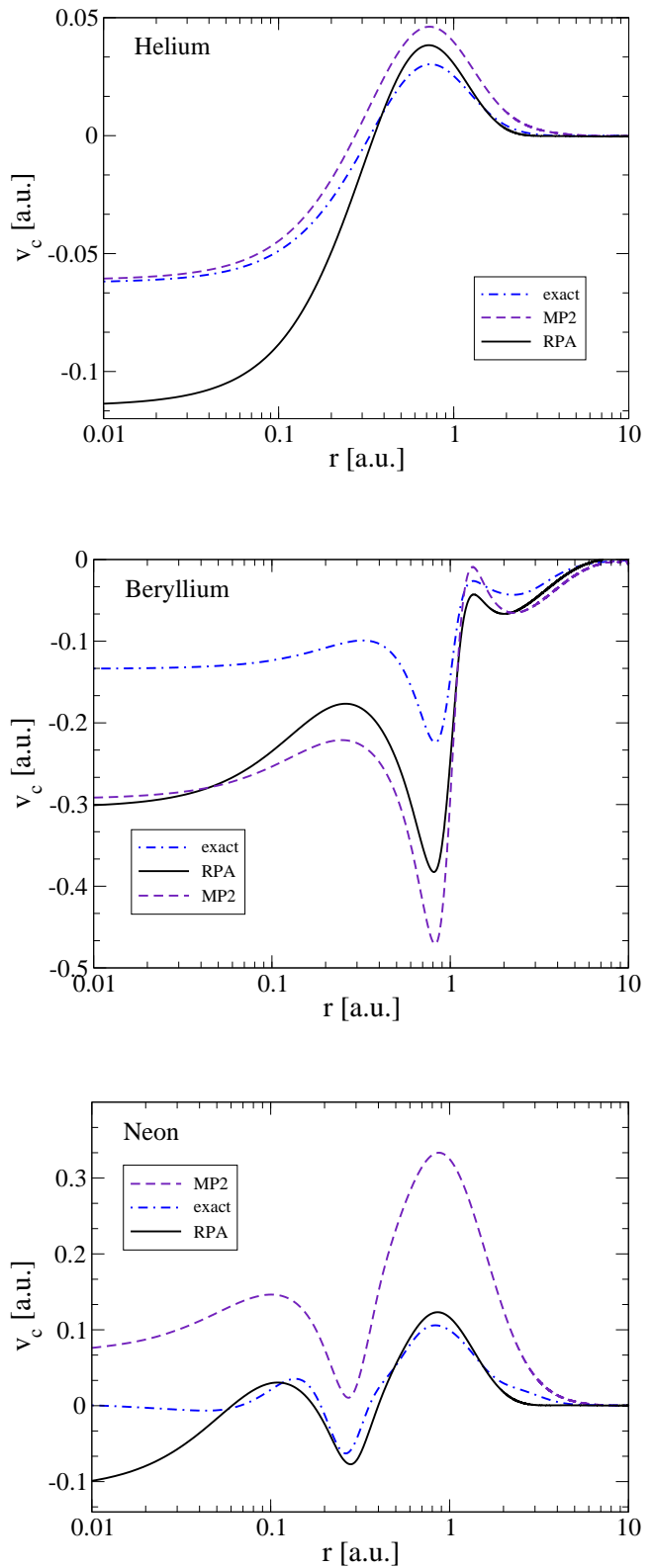

\includegraphics[width=8.5cm, clip=true]{hecorr.eps}\\
\vspace{1cm}
\includegraphics[width=8.5cm,clip=true]{becorr.eps}\\
\vspace{1cm}
\includegraphics[width=8.5cm,clip=true]{necorr.eps}\\
\caption{Self-consistent correlation potentials for He, Be, and Ne. The 
RPA-potential is compared to the MP2-potential and to the exact potential. For 
Be no self-consistent MP2-potential can be obtained. Instead, the potential was evaluated  at the EXX density. For a further discussion see Ref. \onlinecite{engeljiang}}
\label{corrpot}
\end{figure} 
\begin{figure*}
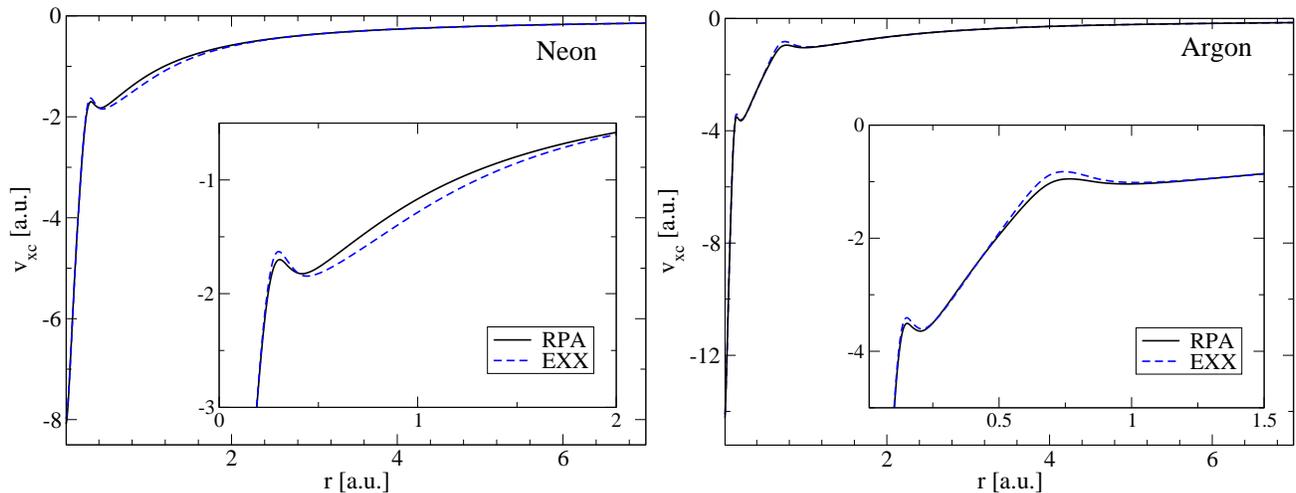

\includegraphics[width=8.5cm, clip=true]{neon.eps}
\includegraphics[width=8.5cm, clip=true]{argon.eps}\\
\caption{The total XC potential $v^{\rm RPA}_{\xc}$ of Ne and Ar compared to $v_{\rm x}$. The effect of correlations is seen to be relatively small and, as 
expected, the shell oscillations are damped by the inclusion of correlation effects.}
\label{armg}
\end{figure*}
\section{Results for spherical atoms}
In the present paper we present results for the spherical atoms He, Be, Ne, Mg and Ar, see Figs. \ref{corrpot}-\ref{armgcorr}. In the cases of He, Be and Ne we compare our results to existing exact density functional potentials.\cite{umr}
\subsection{The RPA correlation potential}
The correlation potentials in this work are defined as the difference between the self-consistent XC potentials calculated within the RPA, or MP2, and that of the 
EXX, 
\be
v_{\rm c}(r)\equiv v_{\xc}[n_{\xc}](r)-v_{\rm x}[n_{\rm x}](r).
\ee
Note that the correlation potential is defined as the difference between potentials 
calculated at two different densities. For a more detailed discussion of this point we refer to Sec. \ref{selfc}. 

In Fig. \ref{corrpot} the RPA and MP2 correlation potentials for He, Be and Ne are presented and compared to the exact correlation potentials.\cite{umr}
The characteristic shell oscillations inherent in the exact potential can not be reproduced by potentials depending explicitly on the density. Indeed, both LDA- and gradient corrected potentials lack the correct shell oscillations (see {\em e.g.} Ref. \onlinecite{engeljiang}). The potential $v_{c}^{\rm RPA}$, however, which is an implicit density functional through the dependence on the KS orbitals and eigenvalues, can be seen to reproduce these oscillations very well and this is also the case for the potential of MP2. Except at the origin, the amplitudes of the oscillations of $v_{c}^{\rm RPA}$ are much closer to those of the exact potential as 
compared to the case of $v_{c}^{\rm MP2}$. 
At the origin the RPA potential appears to deviate the most from the exact potential by being too attractive for all atoms. It should then be remembered that the values of the potential in this region is expected to be of less importance due to the strong singular 
Coulomb potential from the nucleus. It is tempting to interpret the superiority of the RPA correlation potential to that of the MP2 in the outer parts of the atoms as an increased importance of screening in this region. Similarly one might guess that short range correlations are more important in the interior of the atoms leading to a better performance of the MP2 potential in this region.

In order to demonstrate the effect of correlations on the total XC potential we compare the RPA version of $v_{\xc}$ to the potential of $v_{\rm x}$ of EXX in Fig. \ref{armg} (for Ne and Ar). As expected, the shell oscillations are damped by correlation. Due to the good performance of our numerical method the $1/r$ asymptotic behavior can be observed to a large radius without any fitting procedure. According to Niquet 
{\em et al.}\cite{nfgas} the RPA correlation potential should have a polarization tail of the form $-\a^{\rm RPA}_N/2r^4$, where $\a_N$ is the static polarizability of the corresponding atom. This effect can, however, only be observed at 
a considerable distance from the atom, $\approx$ 20 bohr radii. An accurate description of the potential at such large distances requires many more basis functions as compared to our standard calculations. Increasing the accuracy in the asymptotic region, a tendency toward  such a tail was, however, observed. 

\begin{figure}[b]
\includegraphics[width=8.5cm, clip=true]{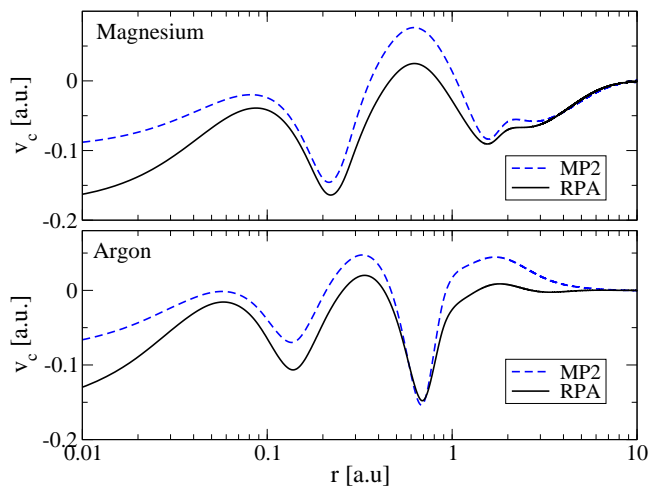}\\
\caption{The RPA and MP2 correlation potentials of Mg and Ar. }
\label{armgcorr}
\end{figure}

Finally, in Fig. \ref{armgcorr} the RPA and MP2 correlation potentials of Mg and Ar are displayed. The presence of a third s-shell, as well as a second p-shell for Ar, can be observed.  
\subsection{Ionization potentials}
Within exact DFT the highest occupied orbital eigenvalue equals the negative of the 
ionization potential.\cite{ab} A further test of the quality of the RPA XC potential is thus a comparison between the eigenenergy of the HOMO and the experimental ionization potential. Results are presented in Table \ref{ioniz}. 
In this table we also present the corresponding results obtained from the EXX and from MP2. The latter approximations also give the proper $-1/r$ tail of the potential which is very important for obtaining a reasonable ionization potential. Therefore, we have not considered it worthwhile to include results from other 
local potentials like, {\em e.g.}, those of the LDA or different GGA:s. As can be 
seen, the ionization potentials of the RPA potential are in very good agreement 
with experiment and represent a significant improvement on the EXX and also on MP2. Consequently, in this regard, the $v_{\xc}^{\rm RPA}$ is the best performing potential presently known. 
\begin{table}[t]
\caption{Ionization potentials for some atoms. The RPA potential produces  very accurate ionization potentials compared to other potentials with the correct $1/r$ decay. Values are in Hartrees.}
\begin{ruledtabular}
\begin{tabular}{cllll}
Atom & RPA & MP2 & EXX &  Exp\footnotemark[2] \\
\hline
He& 0.902 & 0.893 & 0.918 & 0.903  \\
Be& 0.354 & 0.357\footnotemark[1]& 0.309 & 0.343 \\
Ne& 0.796 & 0.657 & 0.851 & 0.792  \\
Mg& 0.297 & 0.303 & 0.253 & 0.281 \\
Ar& 0.590 & 0.560 & 0.591 & 0.579 \\\hline
MAE\footnotemark[3] & 0.009 & 0.040 & 0.030 & 
\label{ioniz}
\end{tabular}
\end{ruledtabular}
\footnotetext[1]{Calculated using the EXX density.}
\footnotetext[2]{Experimental data taken from Ref. \onlinecite{engeljiang}}
\footnotetext[3]{Mean average error.}
\caption{Total ground-state energies calculated from the self-consistent density. The RPA energy functional gives too large correlation energies, whereas the MP2 functional performs much better.}
\begin{ruledtabular}
\begin{tabular}{crrrr}
Atom & RPA & MP2\footnotemark[1] & EXX & Exp\footnotemark[1] \\
\hline
He&2.945&2.909&2.862&2.904\\
Be&14.754&14.697&14.572&14.667\\
Ne&129.143&129.027&128.545&128.937\\
Mg&200.296&200.129&199.612&200.059\\
Ar&527.908 &527.661&526.650&527.604
\label{toten}
\end{tabular}
\end{ruledtabular}
\footnotetext[1]{Experimental data and MP2 values are taken from Ref. \onlinecite{engeljiang}.}
\end{table}
\subsection{Total energies and role of self consistency}\label{selfc}
Using the self-consistent KS orbitals and eigenvalues the total energy was calculated from Eq. (\ref{energi}). The results for different atoms and are presented in Table \ref{toten}. We see that the correlation energy is overestimated by almost a factor of two for the small atoms, leading to a too low total energy. This tendency within the RPA has previously been pointed out by several workers.\cite{furcherpa,nedvbatm} The MP2 energy functional is seen to perform much better in this regard.

We have found the Virial Theorem (VT) to be a convenient test of the accuracy of our calculations. Therefore, in Appendix A, this theorem has been proven by us in the case of any conserving density functional approximation. As a matter of fact, in our calculations, the VT is obeyed to within six significant digits. This is a very satisfactory test on the overall accuracy of our calculations. It is important to point out that the VT only holds if the self-consistent 
orbitals and one-electron eigenvalues are used in the evaluation of the energies.

\begin{figure}[b]
\includegraphics[width=8.5cm, clip=true]{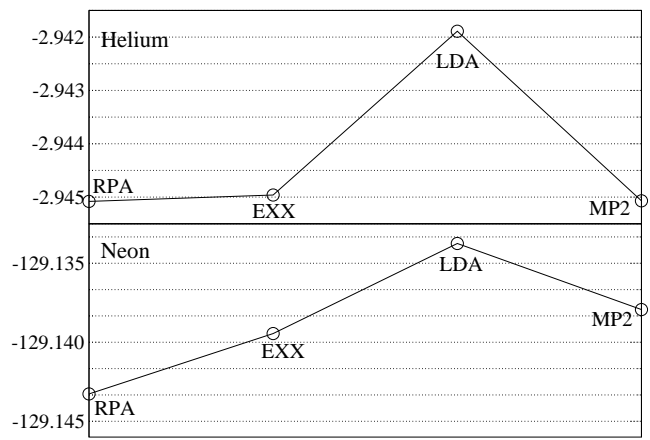}\\
\caption{The total energy for He and Ne in the RPA calculated with orbitals and one-electron eigenvalues from the LDA, EXX, MP2 and the RPA. The RPA orbitals are seen to give the lowest energy confirming that the RPA functional has a minimum.}
\label{min}
\end{figure}

In order to investigate the variational properties of the total energy of the RPA as a function of the density we have evaluated this functional using orbitals and one-electron eigenvalues also from the LDA, MP2 and the EXX. The results are illustrated in Fig. \ref{min}. In all cases we find a higher energy compared to the fully self-consistent RPA result, in accordance with our previously proven theorem that the RPA density functional has a minimum.\cite{nedvlvb}  
We also studied the variation of the potential with respect to the density used for its evaluation. Only small changes in the potential was observed. Comparing to the results of Jiang and Engel\cite{engeljiang} we conclude that the RPA potential is more stable with respect to variations in the density.
\subsection{Two-electron excitation energies}
The particle-conserving or two-electron excitation energies can be obtained 
from the full density response function $\chi$ of the system. In finite systems, 
there are at least a few such excitation energies below the continuum edge which 
then show up as poles of $\chi$. Within exact TDDFT these poles are zeros of the expression $\chi_s^{-1}-v-f_{\xc}$, where $v$ is the bare Coulomb interaction, $f_{\xc}$ is XC kernel, and $\chi_s$ is the non-interacting KS response function with poles at the differences between the DF eigenvalues. In a one-pole approximation, the 
two-particle excitation energies are seen to be the difference between an 
occupied and an unoccupied DF eigenvalue corrected by some matrix elements 
with respect to DF orbitals of the Coulomb interaction and the XC kernel, $f_{\xc}$. 
It has been shown previously by Petersilka {\em et al.}\cite{pgb} that the latter matrix elements have a much smaller influence or effect on the calculated excitation 
energies than the eigenvalue difference obtained by using different approximations to the XC potential. Of course, the presence of a sum over all 
poles will affect the actual zeros of the denominator of $\chi$, {\em i.e.}, the excitation energies. If, however, these zeros are well separated, as is often the case in the discrete part of the spectrum, also this effect is much smaller than the eigenvalue differences produced by different XC potentials. As a consequence, 
an accurate XC potential is of vital importance for obtaining accurate two-particle excitation energies. 

In the Tables \ref{Heexc}-\ref{Neexc} the KS eigenvalue differences in RPA are compared to the exact eigenvalue differences obtained from Ref. \onlinecite{umr} 
and to the MP2 and EXX eigenvalue differences. The results show that the mean average error is significantly reduced as compared to the EXX approximation. For He and Be the MP2 approximation also improves the EXX values. 

The magnitude of the Hartree contribution and the $f_{\xc}$ contribution to the true excitation energies are presented in Tables \ref{Heexc2} and \ref{Beexc2}. We see that the KS eigenvalue differences are already very close to the true excitation energies. Including the Hartree contribution, {\em i.e.}, evaluating the poles of the RPA response function, drives the KS values further away from the true excitation energies. Including the $f_{\xc}$ part cancels this error and in most cases we come back to a value of the same quality as the KS eigenvalue differences. For some transitions, however, {\em e.g.}, the 2s$\rightarrow$2p transition in Be, it is necessary to include both the Hartree and the $f_{\xc}$ contributions.
\begin{table}[t]
\caption{KS eigenvalue differences for He. The values of the RPA and MP2 improves significantly on the EXX values.}
\begin{ruledtabular}
\begin{tabular}{ccccc}
Transition& RPA & MP2 & EXX & Exact\footnotemark[1] \\
\hline
1s$\rightarrow$2s& 0.744 & 0.736 &0.760 & 0.746   \\
1s$\rightarrow$3s& 0.844 & 0.836 &0.861 & 0.839 \\
1s$\rightarrow$2p& 0.775 & 0.768 &0.791 & 0.777 \\
1s$\rightarrow$3p& 0.855 & 0.846 &0.871 & 0.848\\\hline
MAE& 0.004  & 0.006 & 0.018 & 
\label{Heexc}
\end{tabular}
\end{ruledtabular}    
\footnotetext[1]{Taken from Ref. \onlinecite{umr}.}
\caption{KS eigenvalue differences for Be. The errors in the EXX eigenvalue differences are reduces by a factor of two in both the RPA and MP2. Note that the MP2 values are calculated from  the EXX density.}
\begin{ruledtabular}
\begin{tabular}{ccccc}
Transition& RPA & MP2 &EXX & Exact\footnotemark[1] \\
\hline
2s$\rightarrow$3s& 0.254 & 0.253 & 0.214 & 0.244  \\
2s$\rightarrow$2p& 0.131 & 0.128 & 0.131 & 0.133 \\
2s$\rightarrow$3p& 0.276 & 0.276 & 0.234 & 0.269  \\
2s$\rightarrow$4p& 0.332 & 0.330 & 0.297 & 0.305 \\
2s$\rightarrow$3d& 0.292 & 0.292 & 0.241 & 0.283  \\\hline
MAE& 0.011  & 0.011 & 0.023 &  
\label{Beexc}
\end{tabular}
\end{ruledtabular}
\footnotetext[1]{Taken from Ref. \onlinecite{umr}.}
\caption{KS eigenvalue differences for Ne. The RPA results are almost an order of magnitude 
better than those of the EXX which are better than those of MP2.}
\begin{ruledtabular}
\begin{tabular}{crrrr}
Transition& RPA & MP2 & EXX & Exact\footnotemark[1] \\
\hline
1s$\rightarrow$3s& 30.639 & 30.591 &  30.628 & 30.633 \\
1s$\rightarrow$3p& 30.715 & 30.652 &  30.706 & 30.706 \\
1s$\rightarrow$3d& 30.773 & 30.699 &  30.766 & 30.759 \\
2s$\rightarrow$3s& 1.462  &  1.336 &  1.526  & 1.469 \\
2s$\rightarrow$3p& 1.538  & 1.398 &  1.604  &  1.542  \\
2s$\rightarrow$3d& 1.595  & 1.444 &  1.664  & 1.595 \\
2p$\rightarrow$3s& 0.607  & 0.492 &  0.659  & 0.612 \\
2p$\rightarrow$3p& 0.683  & 0.553 &  0.737  & 0.684 \\
2p$\rightarrow$3d& 0.740  & 0.600 &  0.797  & 0.738\\\hline
MAE& 0.007  & 0.111 &  0.046 &
\label{Neexc}
\end{tabular}
\end{ruledtabular}
\footnotetext[1]{Taken from Ref. \onlinecite{umr}.}
\end{table}
\section{Conclusions and Discussion}
In the present work we have calculated that local potential which
through a one-particle Schr\"odinger equation generates the orbitals
which minimize the expression for the total energy within the RPA.
The systems studied are spherical atoms. We have found that the
correlation part of this potential - defined as the total RPA potential
minus that corresponding to the EXX - has the effect of softening the
shell structure produced by the potential of the EXX. This shell
structure consist of rapid oscillations of the XC potential between the
atomic shells.

For several spherical atoms, we have compared the spatial dependence of
our RPA potential to that of the exact DF potential defined to be the
potential which, through the Kohn-Sham procedure, yields the exact
electron density of the many-body system. Such exact potentials exist
in the cases of the He, Be, and Ne atoms. We have found that the
RPA potentials are closer to the exact DF potentials than the
corresponding local potentials of MP2 theory and much closer than the
potentials of the LDA or of any of the GGA:s.

It is relatively difficult to judge the quality of a given local
potential from a study of its spatial dependence. Within DFT, the
highest occupied eigenvalue ought to equal the negative of the ionization
potential. And we have found that our RPA potentials perform very well in
this regard, much better than any traditional potential but also better
than the local potential from MP2 theory. Our average error in the
obtained ionization potentials are $\sim$0.24 eV compared to $\sim$1.1 eV in the case
of MP2 theory and $\sim$0.8 eV within the EXX. We interpret this result as being due to the GWA
providing a better description of correlation effects among the more
loosely bound valence electrons being ionized as compared to the case of
MP2.

Another measure of the quality of the calculated RPA potentials can be
obtained from a study of two-electron excitation energies. Within TDDFT,
and using a single-pole approximation, these can be obtained as
differences between DF eigenvalues corrected by relatively small
matrix elements of the bare Coulomb interaction and the XC kernel. The
latter corrections are relatively insensitive to the orbitals used in their
evaluation and on the choice of XC kernel. Due to the accurate DF
potential of the GWA, {\em i.e.}, the RPA potential, the corresponding
eigenvalues are close to exact DF eigenvalues and the resulting
two-electron excitation energies are also very accurate. 

In the present work, we have shown that the traditional RPA follows
from a special choice of variational expression for the total energy
involving the GWA for the functional $\Phi$ from which the electronic
self-energy is obtained as a functional derivative with respect to the
Green function. This guarantees that the density response function
obtained by perturbing the system by an external potential will be
conserving meaning, {\em e.g.}, that it will obey the f-sum rule. Another
consequence is the fact that the ground-state energies obey the
Virial Theorem which is here proven explicitly for any approximation
within TDDFT obtained from the variational approach to MBPT.

The demonstrated high quality of the XC potential within the RPA,
particularly with regard to energy differences, induces strong hopes that
the density response function obtained by perturbing the system will be very 
accurate indeed. But this will be the topic of a future publication.

In the present paper we have introduced a novel way of doing electronic
structure calculations based on cubic splines as radial basis functions.
The original motivation for the introduction of this somewhat unusual
basis set was the desire to circumvent numerical difficulties
associated with the known singularities of the Kohn-Sham non-interacting
density response function. The latter gives no response to a constant
potential (long-wave-length limit) and a very small response to a very
rapidly varying potential (limit of short wave length). To judge from
the high accuracy of our results the splines appear to be ideally suited
to deal with the latter problem.

We also want to stress two more advantages of our numerical technique
based on the cubic splines. i) the proven accuracy of the re-expansion of
a product of two splines in terms of splines guarantees that the matrices
corresponding to two-particle propagators like, {\em e.g.}, response functions,
are of the same sizes as one-particle propagators like, e.g., Green
functions or - for that matter - wave functions. This property is a
clear advantage over more standard basis sets consisting of, e.g, plane
waves or LMTO:s. ii) The secular problem based on splines requires the
handling of sparse matrices for which there exist efficient standard
computer codes. Our nice experience from using the splines on atomic
problems suggests that we ought to implement similar methods also in
molecules and solids.
\begin{table}[t]
\caption{Excitation energies for He. In the first three columns the difference between the poles of the response function, in three different approximations, and the experimental values are presented. Note that the poles of $\chi_s$ are just the KS eigenvalue differences. The last column gives the experimental values taken from Ref. \onlinecite{pgb}. Already the KS eigenvalue differences are close to the experimental results.}
\begin{ruledtabular}
\begin{tabular}{ccccc}
Transition&$(\chi_s)^{-1}$&$(\chi^{\rm RPA})^{-1}$&$(\chi^{\rm PGG})^{-1}$& Exp. \\
\hline
1s$\rightarrow$2s& $-$0.014 & +0.010 & +0.006 & 0.758   \\
1s$\rightarrow$3s& +0.002 & +0.013 & +0.008 &0.843 \\
1s$\rightarrow$2p& $-$0.005 & +0.005 & +0.002 &0.780 \\
1s$\rightarrow$3p& +0.006 & +0.012 & +0.008 &0.849\\\hline
MAE& 0.007  & 0.010 & 0.006&
\label{Heexc2}
\end{tabular}
\end{ruledtabular}
\caption{Excitation energies for Be. The columns presents the same quantities as in Table \ref{Heexc2}. The 2s$\rightarrow$2p KS transition is seen to be responsible for the large MAE in the first column and needs to be corrected by the Hartree and the $f_{\xc}$ term.}
\begin{ruledtabular}
\begin{tabular}{ccccc}
Transition&$(\chi_s)^{-1}$&$(\chi^{\rm RPA})^{-1}$&$(\chi^{\rm PGG})^{-1}$& Exp.\footnotemark[1] \\
\hline
2s$\rightarrow$3s& +0.005 &  +0.017 & +0.010 &0.249  \\
2s$\rightarrow$4s& +0.013 &  +0.027 & +0.014 &0.297 \\
2s$\rightarrow$2p& $-$0.063 &  +0.024 & $-$0.002 &0.194 \\
2s$\rightarrow$3p& +0.002 &  +0.012 & +0.007 &0.274  \\\hline
MAE&  0.020 & 0.020 & 0.008 &
\label{Beexc2}
\end{tabular}
\end{ruledtabular}
\footnotetext[1]{Taken from Ref. \onlinecite{pgb}.}
\end{table}

Work to apply the ideas introduced in the present paper also to
calculate density response functions and physical properties like, {\em e.g.},
polarizabilities, are in progress.
\begin{acknowledgments}
We are grateful to Prof. Umrigar for allowing us to use unpublished results 
from their work on exact XC potentials. We are also indebted to Dr Engel for communicating data from their calculations on the MP2 potential. M. Stankovski has contributed to our understanding of splines and Prof. Almbladh and Dr Kurth have made several interesting remarks and suggestions. This work was supported by the European Community Sixth Framework Network of Excellence NANOQUANTA (NMP4-CT-2004-500198).
\end{acknowledgments}

\appendix
\section{The Virial Theorem}
We will prove that the Virial Theorem holds for any conserving approximation within DFT generated from the Klein functional. In the full many body case the 
Virial Theorem has already been proved in a conserving approximation.\cite{mp2vir}

Consider the Klein functional as a functional of the density $n$:
\be
Y_{\rm K}[n]= -i \F[G_{s}] + T_s[n] + \int w n + U_{\rm o}.
\ee
Keeping the normalization, the density is scaled with respect to the spatial coordinates,
\be
n^{\lambda}=\lambda^{3}n(\lambda\br).
\label{tat}
\ee 
Due to the stationary property of $Y_{\rm K}$ we have that
\be
\left(\frac{dY_K[n^{\lambda}]}{d\lambda}\right)_{\l=1}=0.
\ee
Let us see how each term in the Klein functional scales when we scale the density as in Eq. (\ref{tat}). The Hartree term scales linearly $(v=1/r)$
\bea
U_{\rm o}^{\lambda}&=&1/2\int d^3rd^3r'v(\br-\br')\lambda^3n(\lambda\br)\lambda^3n(\lambda\br')\nn\\
&=&\l/2\int d^3rd^3r'v(\br-\br')n(\br)n(\br')\nn\\
&=&\lambda U_{\rm o},  
\eea
so
$$
\left(\frac{dU^{\l}_{\rm o}}{d\lambda}\right)_{\l=1}=U_{\rm o}.
$$
The kinetic energy of independent particles is a sum of the occupied one-particle kinetic energies, and is thus an implicit functional of the density. With the density scaled as in Eq. (\ref{tat}) the orbitals scale like
\be
\varphi^{\lambda}=\lambda^{3/2}\varphi(\lambda\br).
\label{orb}
\ee 
Inserting the scaled orbitals in the expression for the kinetic energy we find
\bea
T_s^{\lambda}&=&-\frac{1}{2}\sum_i^{\rm occ}\int d^3r \lambda^{3/2}\varphi_i(\lambda\br)\nabla^2\lambda^{3/2}\varphi_i(\lambda\br)\nn\\
&=&-\frac{1}{2}\sum_i^{\rm occ}\int d^3r \lambda^{2}\varphi_i(\br)\nabla^2\varphi_i(\br)\nn\\ 
&=&\lambda^2 T_{s},  
\eea
and
$$
\left(\frac{dT^{\l}_{s}}{d\lambda}\right)_{\l=1}=2T_{s}.
$$
The external potential energy $W$ scales as
\bea
W^{\lambda}&=&\int d^3r \lambda^3n(\lambda\br)w(\br) \nn\\
&=&\int d^3r n(\br)w(\br/\lambda).   
\eea 
If the external potential is Coulombic we have
$$
\left(\frac{dW^{\l}}{d\lambda}\right)_{\l=1}=W.
$$ 
The XC energy $E_{\xc}$ is not an explicit functional of $n$ but of $G_s$. 
To see how $E_{\xc}$ scales we must first determine how $G_s$ scales, that is, 
to find that Green function $G^{\l}_s$, which corresponds to $n^{\l}$ for every $\l$. Taking into account that the orbitals scale as in Eq. (\ref{orb}) we have$$
-\frac{1}{2}\nabla^2\varphi^{\lambda}_k(\lambda \vr)=-\frac{\lambda^2}{2}\nabla_{\lambda}^2\varphi^{\lambda}_k(\lambda \vr).
$$
From the KS equation, 
$$
\{-\frac{1}{2}\nabla^2+V(\vr)\}\varphi_k(\vr)=\varepsilon_k\varphi_k(\vr),
$$
we see that
$$
-\frac{\lambda^2}{2}\nabla_{\lambda}^2\varphi^{\lambda}_k(\lambda \vr)=-\frac{\lambda^2}{2}\{\varepsilon_k-V(\lambda \vr)\}\varphi^{\lambda}_k(\lambda \vr).
$$
Thus, the equation that yields the scaled orbitals is
\be
\{-\frac{1}{2}\nabla^2+\lambda^2 V(\lambda \vr)\}\varphi^{\lambda}_k(\lambda \vr)=\lambda^2\varepsilon_k\varphi^{\lambda}_k(\lambda \vr).
\ee
Consequently, the Green function scales as
\be
G_s^{\lambda}(\br,\br',\w)=\lambda G(\lambda\br,\lambda\br',\omega/\lambda^2).
\ee
A general $\F$-diagram of order $n$ can be written
$$
\F_n=\frac{1}{2n}\Tr\Sigma_n[G_s]G_s, 
$$
where $\S_n$ is a skeleton diagram of order $n$. In order $n$ there are thus $n$ interaction lines, $2n$ coordinates and $2n$ propagators. Inserting the scaled Green function we find that $\F_n$ scales as 
\be
\F_n^{\l}=\frac{\l^{2-n}}{2n}\Tr\Sigma_n[G_s^{\l}]G_s^{\l}.
\ee
A factor of $\l^{2n}$ comes from the number of propagators. The variable substitution gives us a factor of $\l^{-6n}$ and from the interaction lines an additional factor of $\l^n$ is obtained. There are $n+1$ $\w$-integrations giving a factor of $\l^{2(n+1)}$. In total we obtain a factor of $\l^{2-n}$. Taking the $\l$-derivative we obtain
$$
\left(\frac{d\F_n^{\l}}{d\lambda}\right)_{\l=1}=(2-n)\F.
$$
Summing over $n$ yields
\bea
\left(\frac{d\F^{\l}}{d\lambda}\right)_{\l=1}&=&\sum_n\left(\frac{d\F_n^{\l}}{d\lambda}\right)\nn\\
&=&\sum_n(2-n)\F_n\nn\\
&=&2\F-\sum_nn\F_n\nn\\
&=&2iE_{\xc}[G_s]-iU_{\xc}[G_s].
\eea
where
$$
U_{\xc}[G_s]=-\frac{i}{2}\sum_n\Tr\Sigma^n[G_s]G_s.
$$
Summing all the terms we have proved the Virial Theorem
\bea
0&=&2T_s+U_{\rm H}+W+2E_{\xc}-U_{\xc}\nn\\
&=&2T_s+2T_{\xc}+U_{\rm H}+W+U_{\xc},
\eea
where
$$
T_{\xc}\equiv E_{\xc}-U_{\xc}.
$$


\begin{thebibliography}{46}
\bibitem[{txt()}]{txt}
\bibinfo{note}{It should, however, be remembered that their work was based on
  pseudo potentials thus excluding the possibility of treating the effect of
  core-valence exchange in a proper way. The numerical consequences of this is
  still an open question.}
\end{thebibliography}
\end{document}